\documentclass[11pt,twoside]{article}

\usepackage{asp2004}
\usepackage{epsf}
\usepackage{psfig}

\begin{document}
\title{Studying distant dwarf galaxies with GEMS and SDSS}   
\author{Fabio D. Barazza$^1$, Shardha Jogee$^1$, Hans-Walter Rix$^2$, Marco
Barden$^2$, Eric F. Bell$^2$, John A. R. Caldwell$^3$, Daniel H. McIntosh$^4$,
Klaus Meisenheimer$^2$, Chien Y. Peng$^5$, Christian Wolf$^6$}   
\affil{$^1$Department of Astronomy, University of Texas at Austin, Austin, USA,
$^2$Max-Planck Institute for Astronomy, Heidelberg, Germany, $^3$McDonald
Observatory, University of Texas, Fort Davis, USA, $^4$Department of Astronomy,
University of Massachusetts, Amherst, USA, $^5$Space Telescope Science
Institute, Baltimore, USA, $^6$Astrophysics, University of Oxford, Oxford,
U.K.}    

\begin{abstract} 
We study the colors, structural properties, and star formation histories of a
sample of $\sim1600$ dwarfs over look-back times of $\sim3$ Gyr
($z=0.002-0.25$). The sample consists of 401 distant dwarfs drawn from the
Galaxy Evolution from Morphologies and SEDs (GEMS) survey, which provides high
resolution {\it Hubble Space Telescope (HST)} Advanced Camera for Surveys (ACS)
images and accurate redshifts, and of 1291 dwarfs at 10--90 Mpc compiled from
the Sloan Digitized Sky Survey (SDSS). We find that the GEMS dwarfs are bluer
than the SDSS dwarfs, which is consistent with star formation histories
involving starbursts and periods of continuous star formation. The full range
of colors cannot be reproduced by single starbursts or constant star formation
alone. We derive the star formation rates of the GEMS dwarfs and estimate the
mechanical luminosities needed for a complete removal of their gas. We find
that a large fraction of luminous dwarfs are likely to retain their gas,
whereas fainter dwarfs are susceptible to a significant gas loss, {\it if} they
would experience a starburst.
\end{abstract}

\section{Introduction}
The evolution of dwarfs is a complex problem, where evolutionary paths may
depend on a variety of external and internal factors. Our knowledge of the
local volume ($<8$ Mpc) has deepened, in particular due to strong efforts in
determining distances to many nearby galaxies \citep[and references
therein]{kar03}. However, it is still unclear what governs the evolution of
dwarfs in low density regions and how the different morphological types form.

Here, we present a study of the properties of dwarf galaxies over the last 3
Gyr ($z=0.002-0.25$)\footnote{We assume a flat cosmology with
$\Omega_M = 1 - \Omega_{\Lambda} = 0.3$ and
$H_{\rm 0}$=70~km~s$^{-1}$~Mpc$^{-1}$.} drawn from GEMS \citep{rix04} and SDSS
\citep{aba04}.

\section{Basic properties of the sample}
\begin{figure}
\plottwo{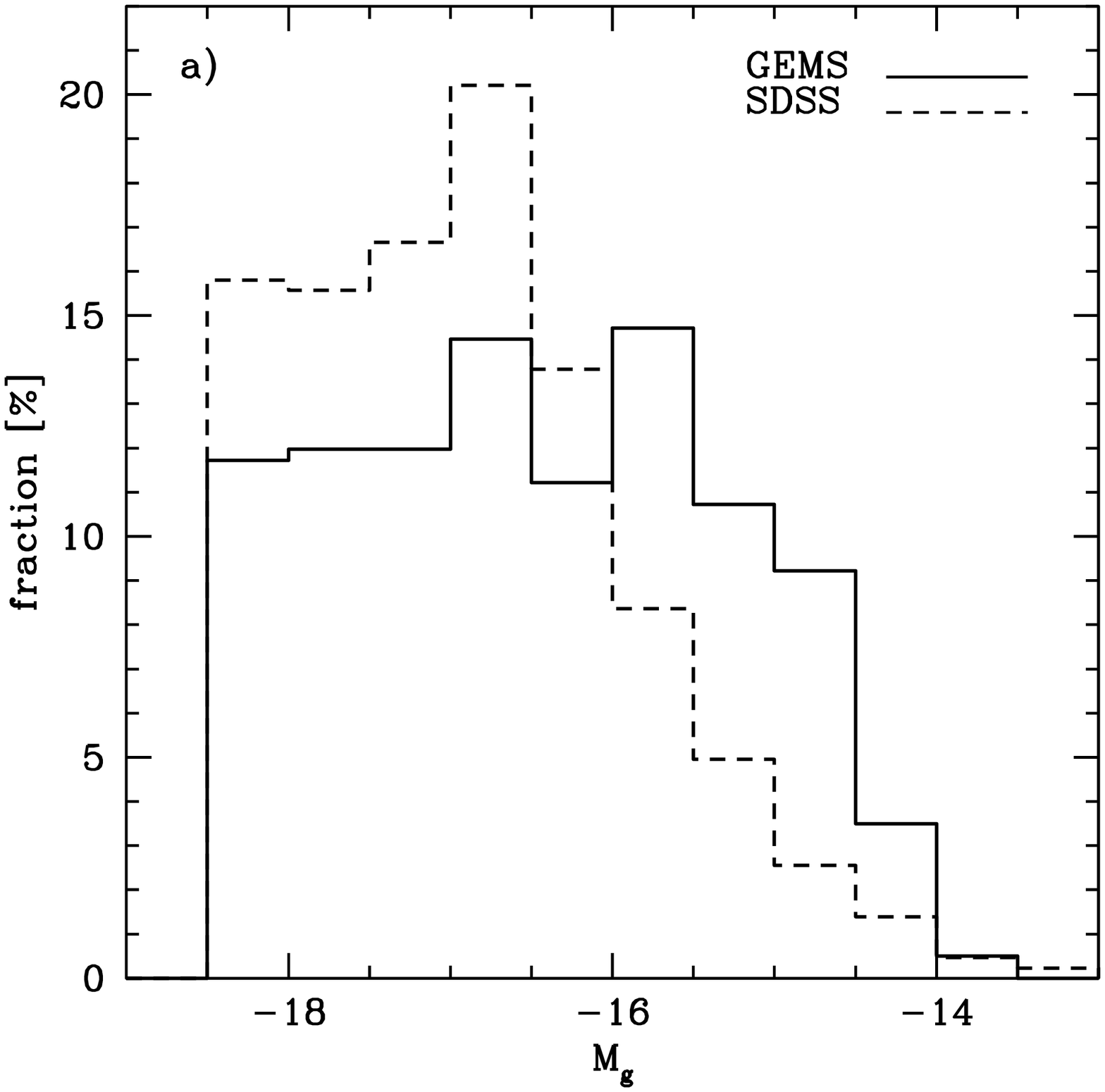}{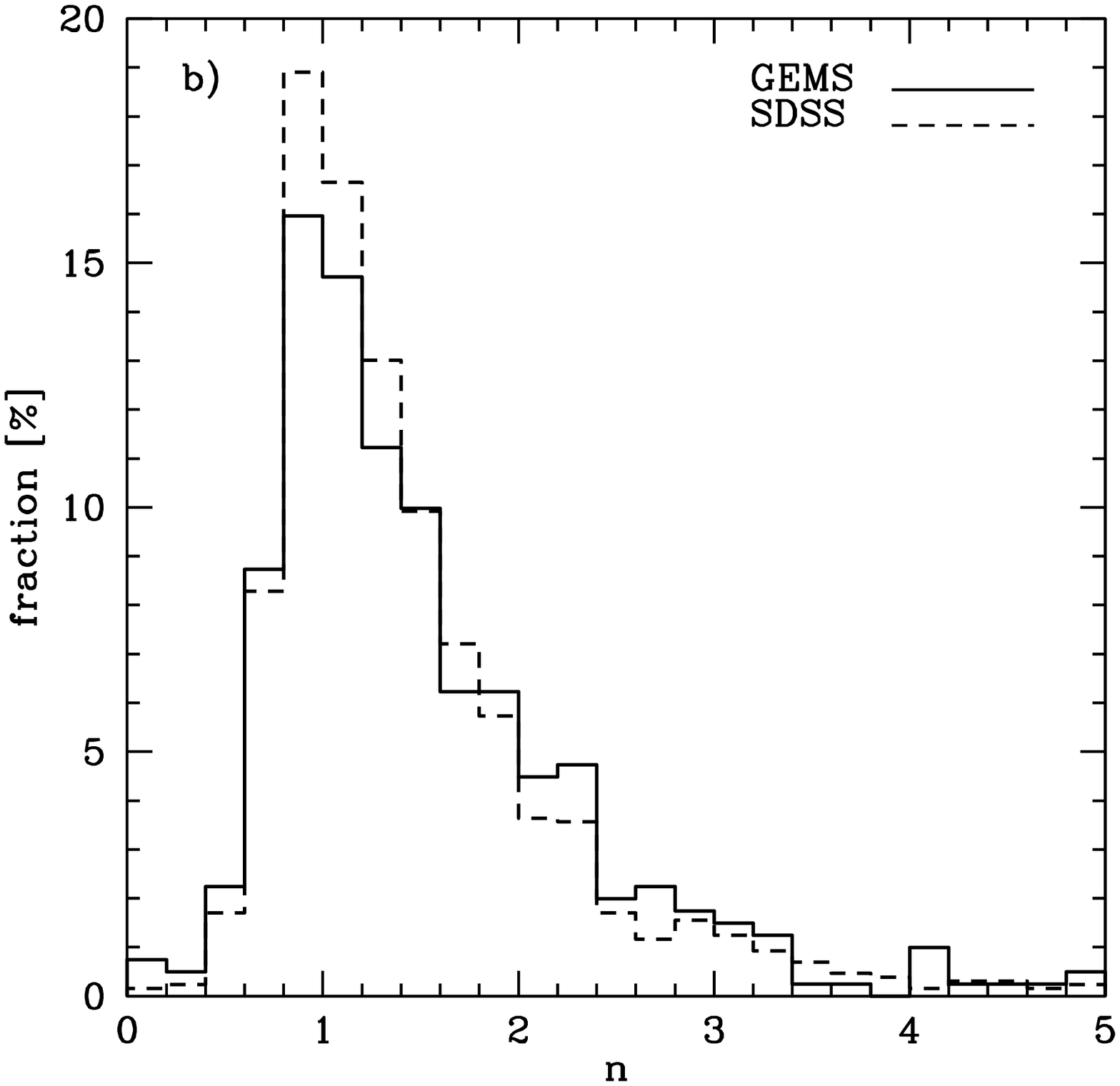}
\caption{{\bf a)} The $g$-band luminosity distribution of the GEMS and SDSS
samples. The median value for the GEMS dwarf sample is $-16.51$ mag and for
SDSS $-16.95$ mag. {\bf b)} Histograms of the Sersic index $n$. The median
values are 1.32 and 1.25 for GEMS and SDSS, respectively.\label{f1}}
\end{figure}

Our starting sample consists of 988 dwarfs from the GEMS survey in the redshift
range $z\sim0.09-0.25$ (corresponding to look-back times of 1 to 3 Gyr), and a
comparison local sample of 2847 dwarfs with $z<0.02$ from the NYU Value-Added
low-redshift Galaxy Catalog \citep[NYU-VAGC,][]{bla05} of the SDSS, which have
been identified and extracted by applying an absolute magnitude cut of
$M_g>-18.5$ mag. The surface brightness profiles of the dwarfs in this sample
have subsequently been fitted with a Sersic model using GALFIT \citep{pen02}.
Finally, we limited the sample to objects with an effective surface brightness
brighter than 22 mag arcsec$^{-2}$ in the $z$ band, which corresponds to the
completeness limit of the SDSS sample \citep{bla05}. The final sample consists
of 401 dwarfs from GEMS and 1291 dwarfs from SDSS. Figure \ref{f1} shows the
distributions of the luminosities and Sersic indices.

\section{Global colors and star formation histories}
\begin{figure}
\plottwo{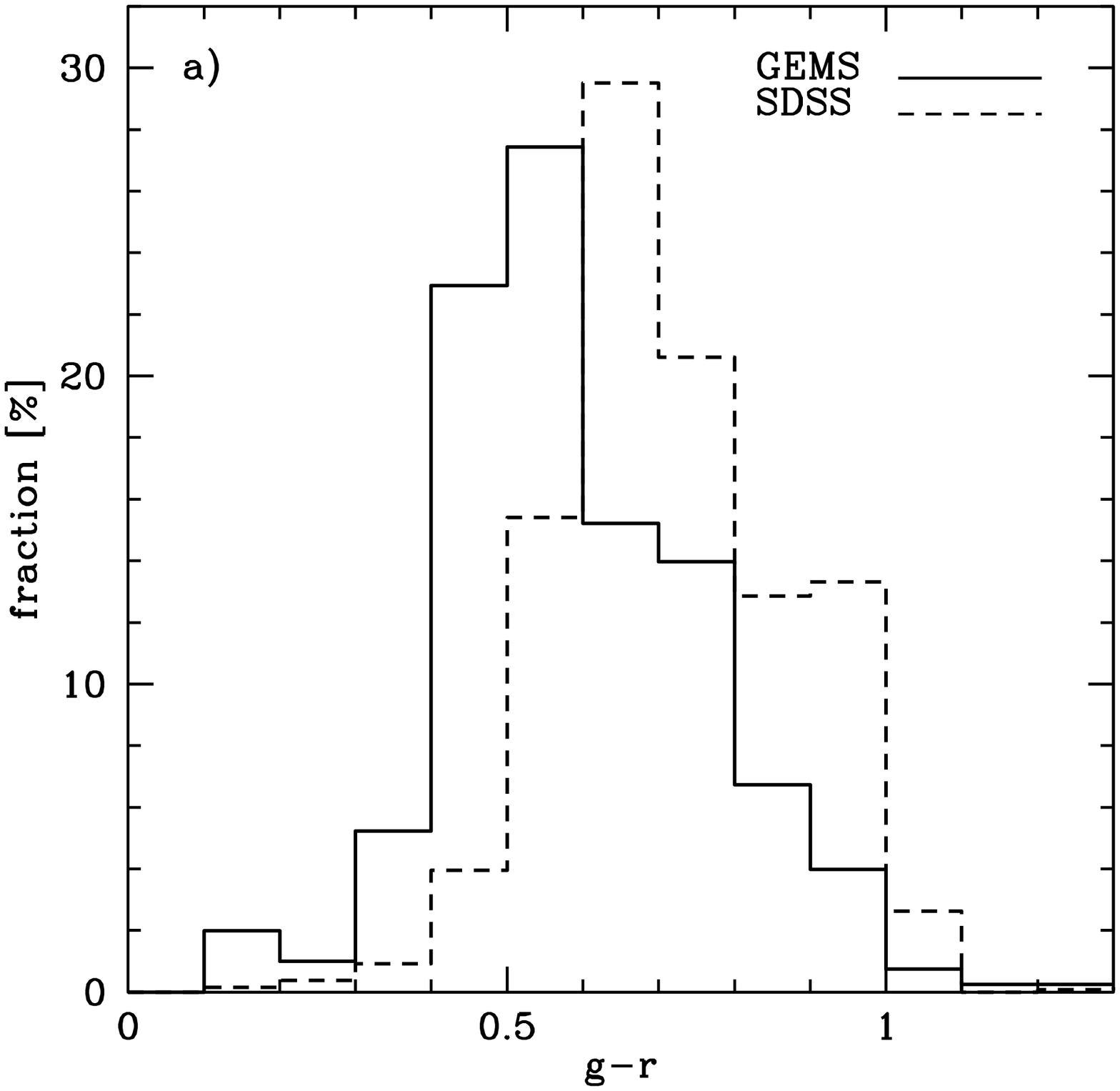}{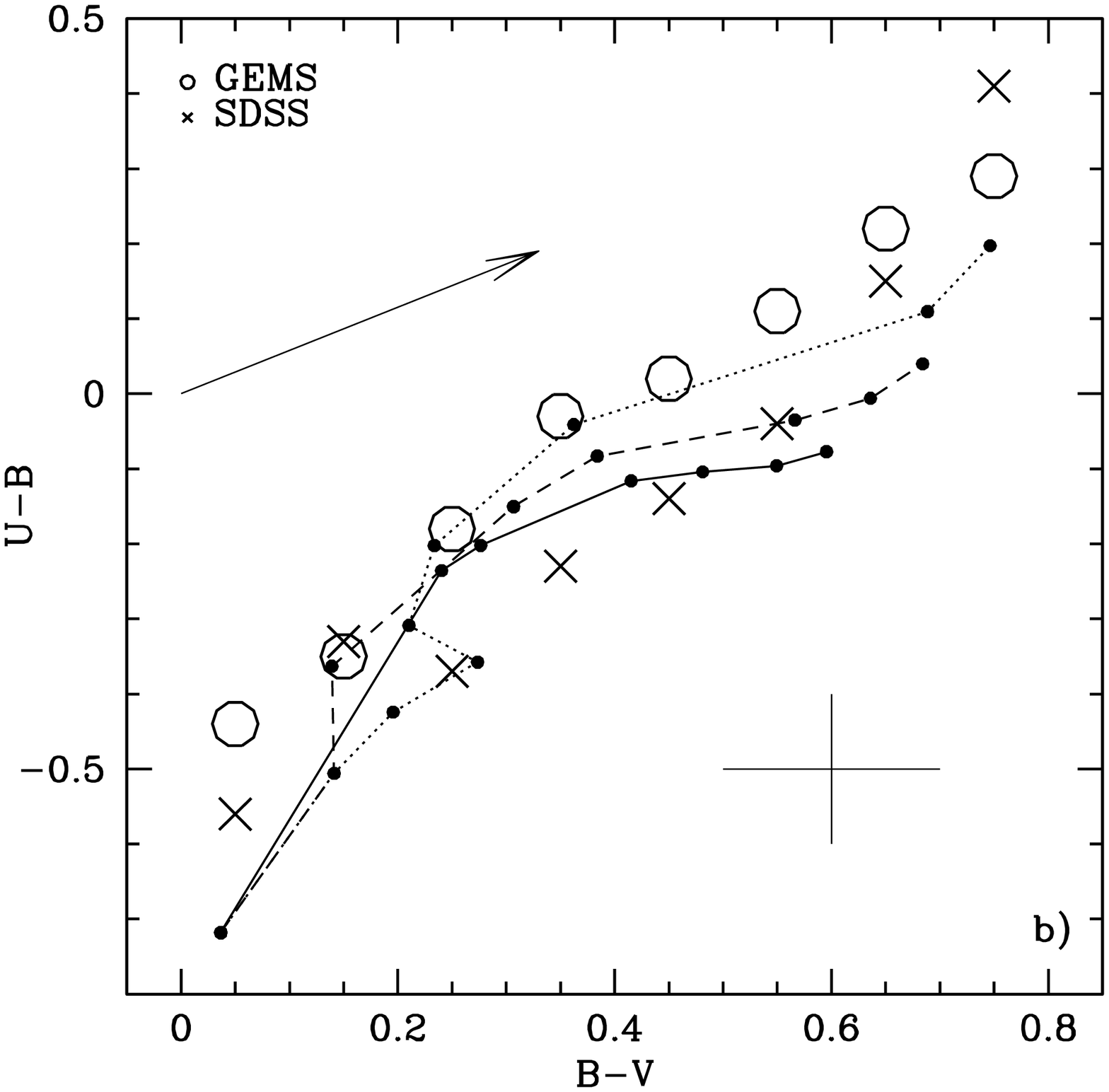}
\caption{{\bf a)} The $g-r$ color distributions for both samples. The median
colors are 0.57 and 0.70 for GEMS and SDSS, respectively. {\bf b)} Color-color
plot for the two samples. The distribution of the galaxies is represented by
the mean $U-B$ colors in 0.1 mag $B-V$ color bins. The lines represent models,
where a continuous SF with a constant rate of $SFR=0.03$ M$_{\sun}$ yr$^{-1}$
and a metallicity of $Z=0.004$ has been combined with various single SBs
starting at different times. For all models a Kroupa IMF has been used. Models
start at 0.1 Gyr (left) and end at 15 Gyr (right) {\it Solid line:} A single SB
with a mass of $3\times10^8$ M$_{\sun}$ and $Z=0.0004$ starts at 0.1 Gyr.
{\it Dashed line:} A single SB with a mass of $3\times10^8$ M$_{\sun}$ and
$Z=0.004$ starts at 0.9 Gyr. {\it Dotted line:} A single SB with a mass of
$5\times10^8$ M$_{\sun}$ and $Z=0.02$ starts at 3.9 Gyr. The error bars
represent the errors of the single color measurements. The arrow indicates the
effect of dust on the colors \citep{sch98}.\label{f2}}
\end{figure}

The comparison of the global colors of the two samples shows that the GEMS
dwarfs are significantly bluer than the SDSS dwarfs, which is apparent in the
histograms shown in Figure \ref{f2}a. A KS-test yields a probability of
$\sim2\times10^{-41}$ that the two color distributions stem from the same
parent distribution. In order to examine the origin of this color difference,
we compare the colors of our sample dwarfs to star formation (SF) models from
{\it Starburst 99} \citep{lei99} in Figure \ref{f2}b. The general color
difference between the two samples is consistent with the color evolution of
models combining a starburst (SB) with continuous SF on a low level. The model
tracks show rather long periods of time with roughly constant $U-B$ colors,
while the $B-V$ colors are becoming significantly redder. This reddening occurs
over time spans comparable to the average look-back time of the GEMS sample and
the amount of reddening is in good agreement with the color difference between
the two samples.

\section{Star formation rates and feedback}
\begin{figure}
\plottwo{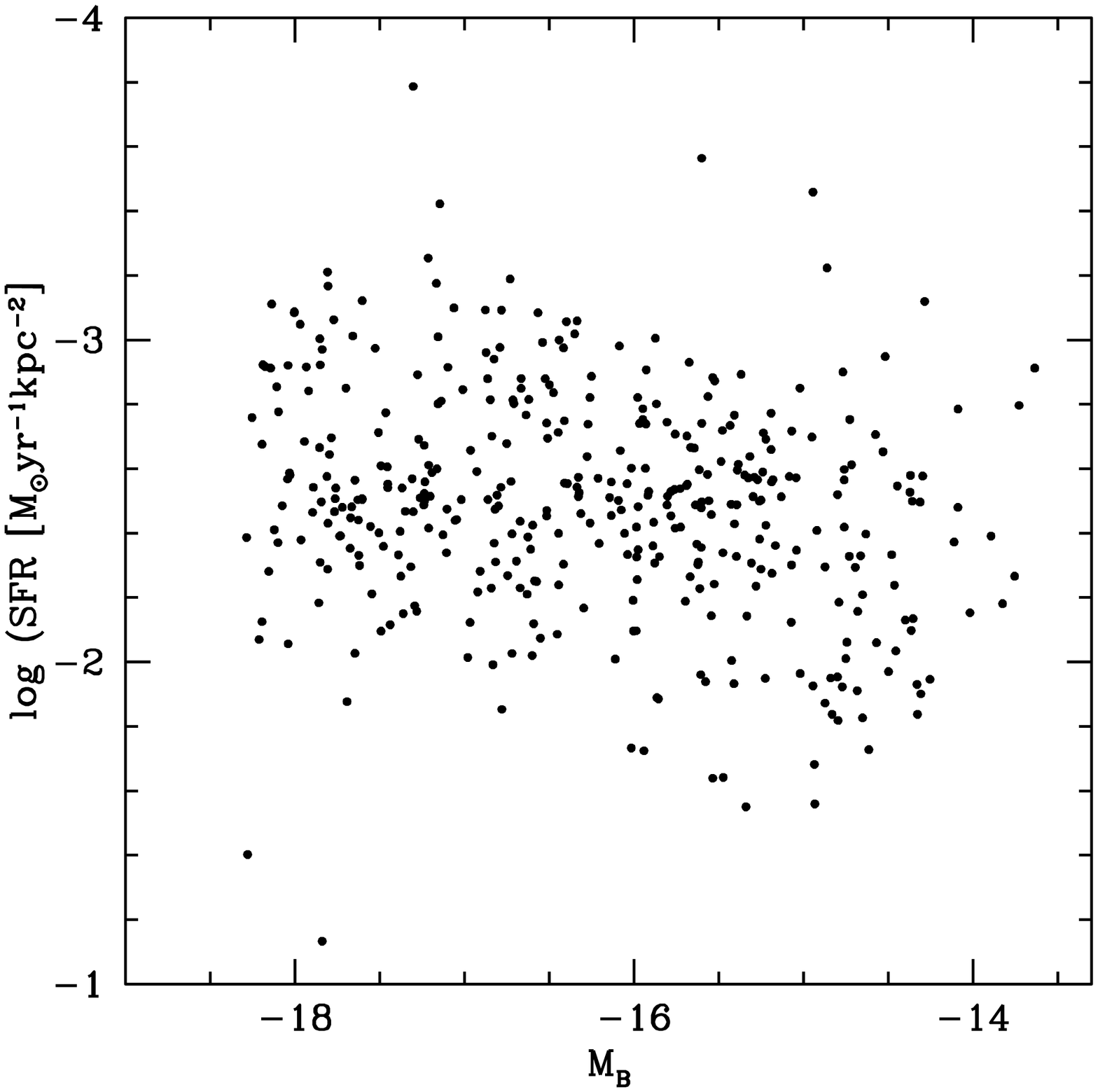}{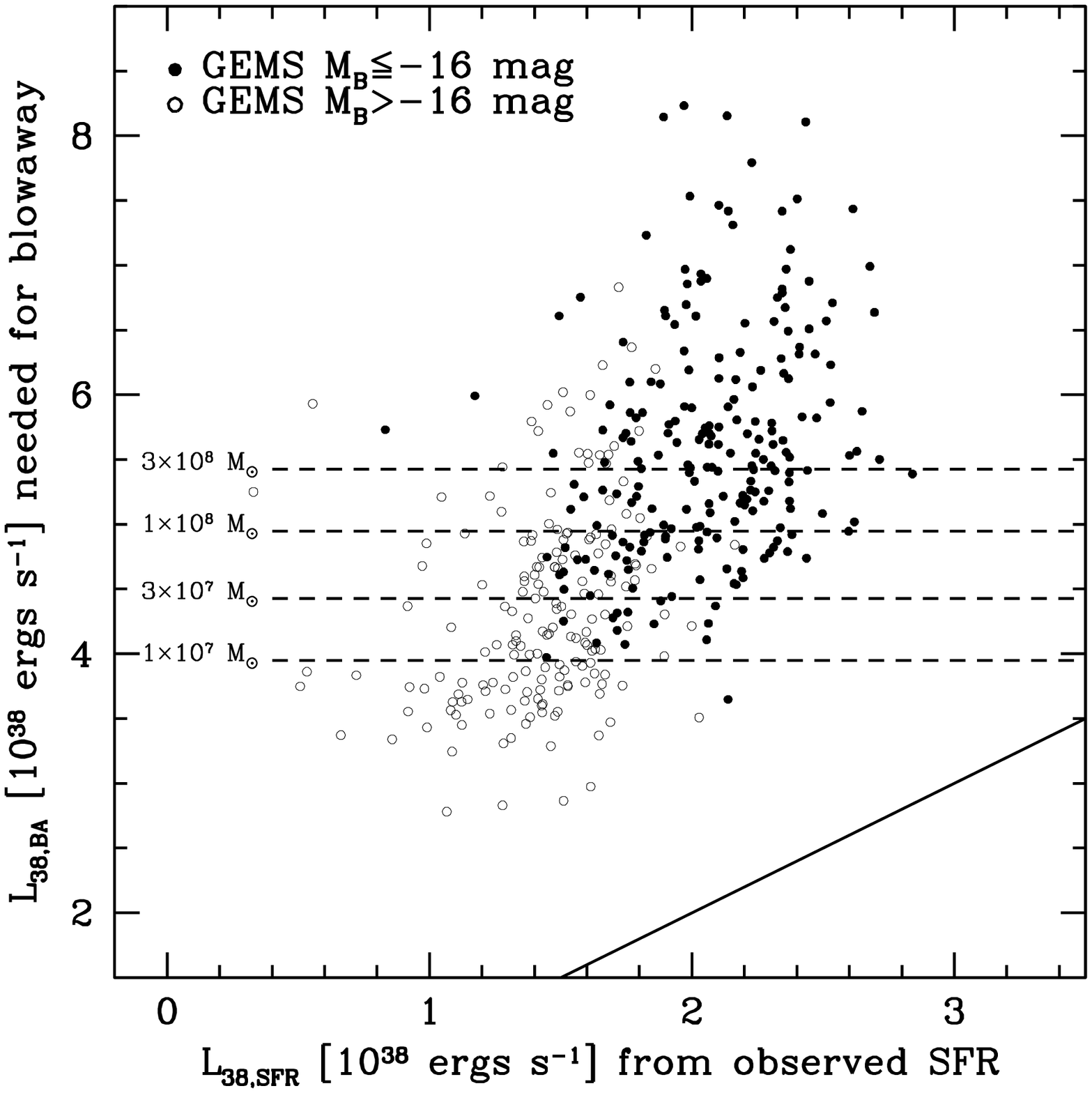}
\caption{{\bf a)} The normalized SFR versus $M_B$. The SFRs have been estimated
from the 2800\AA~continuum fluxes ($L_{2800}$) and using the equation
$SFR$[M$_{\sun}$yr$^{-1}]=3.66\times10^{-40}L_{2800}$ [ergs
s$^{-1}~\lambda^{-1}]$ adopted from \citet{ken98}. These SFRs have then been
divided by the isophotal area provided by Sextractor. {\bf b)} Plot of the ML
needed for a complete blowaway of the gas in dwarfs versus the MLs inferred
from the SFRs. The four dashed lines mark the peak MLs reached of SBs with the
indicated masses. The solid line corresponds to
$L_{38,SFR}=L_{38,BA}$.\label{f3}}
\end{figure}

Using the rest-frame luminosity in a synthetic UV band centered on the
2800\AA~line, we estimate the star formation rate (SFR) of the dwarfs in the
GEMS sample, assuming continuous SF over the last $10^8$ years, which is likely
the case for a majority of our dwarfs. In Figure \ref{f3}a we plot the {\it
normalized} SFR versus $M_B$. In a next step, we estimate the mechanical
luminosities (MLs) needed for the complete removal of the gas from the dwarfs.
The estimate is based on the blowaway model by \citet{mac99}. In this model,
the  MLs depend only on the mass, which we derive from the $V$-band
luminosities, and the ellipticity. In Figure \ref{f3}b we compare these MLs
with the ones expected for the measured SFRs. We find that for their derived SF
histories, the luminous ($M_B=-18$ to $-16$ mag) dwarfs are likely to retain
their gas and avoid blowaway. However, there are a fair number of low
luminosity dwarfs ($M_B=-14$ to $-16$) that are susceptible to a complete
blowaway of gas, {\it if they were} to experience a SB. However, in practice,
only a small fraction of these low luminosity dwarfs {\it may be actually
undergoing} a SB. Even though, we do not have any clear evidence that some
dwarfs in our sample experience a SB at the time of observation, we are also
not able to rule this out. The derived MLs stem from the SFRs, which have been
determined assuming that the dwarfs had a constant SFR over the last $10^8$
years. In addition, we used the near-UV luminosities, which could be affected
by dust. In view of these uncertainties, the derived SFRs have to be considered
as lower limits.

\acknowledgements
F.D.B. and S.J. acknowledge support from the National Aeronautics and Space
Administration (NASA) LTSA grant NAG5-13063 and from HST-GO-10395 and
HST-GO-10428. E.F.B. was supported by the European Community's Human Potential
Program under contract HPRN-CT-2002-00316 (SISCO). C.W. was supported by a
PPARC Advanced Fellowship. D.H.M acknowledges support from the NASA LTSA Grant
NAG5-13102. Support for GEMS was provided by NASA through number GO-9500 from
STScI, which is operated AURA, Inc., for NASA, under NAS5-26555.

\end{document}